\documentclass[slac_one]{revtex4}
\usepackage{graphicx}
\usepackage{fancyhdr}
\begin{document}

\title{{\small{2005 International Linear Collider Workshop - Stanford,
U.S.A.}}\\ 
\vspace{12pt}
Charged Higgs Production in Association with W Boson

at Photon Colliders
\footnote{This talk was given by E.~Asakawa.}
} 

%

\author{E. Asakawa}
\affiliation{YITP, Kyoto University, Kyoto 606-8502, Japan}
\author{O. Brein}
\affiliation{Institute f\"ur Theoretische Physik E,
RWTH Aachen, D-52056 Aachen, Germany}
\author{S. Kanemura}
\affiliation{Department of Physics, Osaka University, Toyonaka,
Osaka 560-0043, Japan}

\begin{abstract}
It is important to explore the Higgs sector 
in order to identify the model beyond the standard model.
We study the charged Higgs production
in the $\gamma\gamma$ mode of a linear collider (LC)
with $1000$ GeV center of mass energy. We show that
the cross sections for the $\gamma \gamma \rightarrow H^\pm W^\mp$
processes can be significantly enhanced
in the two Higgs doublet model (THDM).
The cross section 
can reach $0.1 - 100$ fb which
is comparable to other
charged Higgs boson production processes at a photon collider.
While for other processes the cross sections are too small
for $m_H^\pm \geq 500$ GeV, $0.1 - 100$ fb can be expected
in the $\gamma\gamma \rightarrow H^\pm W^\mp$
processes for $m_H^\pm \geq 570$ GeV when $m_{A^0}=800$ GeV. 
Therefore, 
even if the charged Higgs bosons can not be detected at 
the Large Hadron Collider (LHC), and
even if the charged Higgs bosons 
are too heavy to be detected in other charged Higgs boson
production processes at the LC,
it may be possible to detect them in this process.
\end{abstract}

\maketitle

\thispagestyle{fancy}


\section{INTRODUCTION} 

The Higgs sector of the standard model (SM) can be extended by 
additional Higgs multiplets. In models with additional
isospin $SU(2)$-doublets, 2 neutral and a pair of charged 
Higgs bosons for each additional doublet are newly included
as physical Higgs bosons. 
Then,
searches for the Higgs bosons and measurements of their properties
are indispensable for understanding the mechanism of electroweak
symmetry breaking and also physics beyond the SM.

The decay of a Higgs boson into another Higgs boson
is worthwhile to study, because we can observe two kinds of 
Higgs bosons simultaneously. It is also important in the study of
strategies for the detection of Higgs bosons.
However,
this type of decay
is hard to occur in the decoupling region
of the minimal supersymmetric
standard model (MSSM) because quartic coupling constants 
in the Higgs potential
whose linear combination is related to the mass difference among
heavy Higgs bosons are restricted to
the electroweak gauge coupling constants $g$ and $g'$.
On the other hand, in the general extended Higgs sector, such as 
the THDM, this type of decays can be allowed
within some constraints. Therefore, to study
the decay of a Higgs boson into another Higgs boson
may bring us a opportunity to distinguish between the MSSM
and such a model~\cite{ABK} as well as to affect
detection strategy for Higgs bosons.

In this talk, we concentrate on the decay of a heavy neutral Higgs boson
into a charged Higgs boson in the THDM.
For production of heavy neutral Higgs bosons,
a photon collider can be an ideal device where neutral Higgs bosons
are produced in the $s$-channel via loops of charged particles.
Its reach for the mass of Higgs bosons is about $80\%$ of
the parent $e^+ e^-$ collider.
Even if the charged Higgs bosons can not be detected at the LHC, and even if
the charged Higgs bosons
are too heavy to be detected in other charged Higgs boson
production processes at the LC~\cite{MK,KMO},
it may be possible to detect them in this decay of neutral Higgs bosons
at a photon collider.

\section{THE TWO HIGGS DOUBLET MODEL}
We consider the THDM model with a (softly broken) discrete symmetry
under the transformation $\Phi_1 \rightarrow \Phi_1$ and
$\Phi_2 \rightarrow -\Phi_2$, where $\Phi_i$ are the Higgs isodoublets
with hypercharge $1/2$.
The Higgs potential at the tree level is given by 
\begin{eqnarray}
V(\Phi_1, \Phi_2)&=&m_1^2 |\Phi_1|^2 +m_2^2 |\Phi_2|^2
                 -m_3^2 \Phi_1^\dag \Phi_2 -m_3^{*2}\Phi_2^\dag \Phi_1
\\ \nonumber
              &~&+\frac{\lambda_1}{2} |\Phi_1|^4
                 +\frac{\lambda_2}{2} |\Phi_2|^4
		 +\lambda_3 |\Phi_1|^2 |\Phi_2|^2
                 +\lambda_4 |\Phi_1^\dag \Phi_2|^2
                 +\frac{\lambda_5}{2} (\Phi_1^\dag \Phi_2)^2 
                 +\frac{\lambda_5^*}{2} (\Phi_2^\dag \Phi_1)^2 
\end{eqnarray}
where $m_1^2$, $m_2^2$ and $\lambda_1$ to $\lambda_4$ are real.
Since we consider the case of no CP violation in the Higgs sector, 
$m_3^2$ and $\lambda_5$ are also real though they are generally complex.
The potential has its minimum at $\Phi_i=(0, v_i/\sqrt{2})^T$ ($i=1,2$).
The 8 parameters in the potential can transform into 
the masses of the Higgs bosons, $m_{h^0}$,$m_{H^0}$,$m_{A^0}$ and $m_{H^\pm}$,
the mixing angles $\alpha$ and $\beta$, the vacuum expectation value
$v=\sqrt{v_1^2 +v_2^2}=\sqrt{2^{-\frac{1}{2}}G_F^{-1}}$, and
the soft-breaking scale of the discrete symmetry
$M^2 \equiv m_3^2/(\sin\beta \cos\beta)$.
Using these physical parameters,
the quartic coupling constants $\lambda_1-\lambda_5$ can be written as
\begin{eqnarray}
\lambda_1&=&\frac{1}{v^2 \cos^2\beta}
(-\sin^2 \beta M^2 + \sin^2 \alpha m_{h^0}^2 + \cos^2 \alpha m_{H^0}^2),
\\
\lambda_2&=&\frac{1}{v^2 \sin^2\beta}
(-\cos^2 \beta M^2 + \cos^2 \alpha m_{h^0}^2 + \sin^2 \alpha m_{H^0}^2),
\\
\lambda_3&=&-\frac{M^2}{v^2} + 2\frac{m_{H^\pm}^2}{v^2}
+ \frac{1}{v^2} \frac{\sin 2\alpha}{\sin 2\beta} (m_{H^0}^2-m_{h^0}^2),
\\
\lambda_4&=&\frac{1}{v^2}(M^2 + m_{A^0}^2 - 2m_{H^\pm}^2),
\label{lam4} \\
\lambda_5&=&\frac{1}{v^2}(M^2 - m_{A^0}^2).
\label{lam5}
\end{eqnarray}

\section{The $\gamma \gamma \rightarrow H^\pm W^\mp$ processes}

In the THDM including the MSSM,
the processes $\gamma \gamma \rightarrow H^\pm W^\mp$ 
are loop-induced processes, which consist of
the triangle-type and box-type diagrams in the leading order.
The triangle diagrams lead to the $s$-channel exchanges of
neutral Higgs bosons ($\phi=h^0, H^0, A^0$).
Since the quark contributions in loops tend to cancel between
the triangle-type and box-type diagrams
in the case insensitive to the $s$-channel Higgs resonance~\cite{DHKR},
the resonant $\phi$ production process can dominate around the resonance.
In this talk, we consider the case where the peaked energy distribution
of $\gamma\gamma$ collision is adjusted to mass of a 
neutral Higgs boson followed by decay into a charged Higgs boson
and a $W$ boson.

From Eqs.~(\ref{lam4}) and (\ref{lam5}) we see that the mass splitting
between the CP-odd Higgs boson $A^0$ and 
the charged Higgs boson $H^\pm$ is determined by
$\lambda_4$ and $\lambda_5$ only:
\begin{eqnarray}
m_{A^0}^2 - m_{H^\pm}^2 = \frac{v^2}{2}(\lambda_4 - \lambda_5).
\end{eqnarray}
Let us
consider the $A^0$ resonant production followed by the decay into 
$H^\pm$ and $W^\mp$ in the THDM. It is notable that,
in the MSSM, $\lambda_4=-g^2/2$ and $\lambda_5=0$ lead to
$m_{A^0}^2 - m_{H^\pm}^2 = -m_W^2 $ which can not induce
the resonant $A^0$ production in the 
$\gamma \gamma \rightarrow H^\pm W^\mp$ processes. 

The vertex which induces the $A^0$ decay into $H^\pm$ and $W^\mp$
at the tree level can be written as  
\begin{eqnarray}
{\cal V}_{H^{\pm} W^{\mp} A^0}=\frac{g}{2}(p_{A^0}-p_{H^\pm}),
\end{eqnarray}
where
$p_{A^0}$ and $p_{H^\pm}$ are the momentum of $A^0$ and $H^\pm$
in the in-coming directions to the vertex.
The helicity amplitude for the process
$\gamma(\lambda_1) \gamma(\lambda_2) \rightarrow A^0 \rightarrow
H^{\pm} W^{\mp}$ where we sum over the helicities of the out-going
$W$ boson
is given by the multiplication
of the Higgs-$\gamma\gamma$ vertex function ${\cal A}^{\lambda_1 \lambda_2}$,
the Higgs propagator ${\cal B}$ and the decay part ${\cal C}$:
\begin{eqnarray}
{\cal M}^{\lambda_1 \lambda_2}={\cal A}^{\lambda_1 \lambda_2}
\cdot {\cal B} \cdot {\cal C}.
\end{eqnarray}
Each part can be written as 
\begin{eqnarray}
{\cal A}^{\lambda_1 \lambda_2} &=& \frac{(\lambda_1 + \lambda_2)}{2} 
\frac{\alpha g}{8\pi}
\frac{m_{A^0}^2}{m_W}\sum_{i} I_{A^0}^i,
\\
~~{\cal B}~~ &=& \frac{-1}{s_{\gamma\gamma}-m_{A^0}^2+i m_{A^0} \Gamma_{A^0}},
\\
~~{\cal C}~~ &=& g\frac{m_{A^0}}{m_W}E_W \beta_W,
\end{eqnarray}
where $\lambda_1$ and $\lambda_2$ denote the helicities of the colliding photons.
$\sqrt{s}_{\gamma\gamma}$ and $\Gamma_{A^0}$ are 
the $\gamma\gamma$ collision energy and the total decay width of $A^0$,
and 
the functions $I_{A^0}^i$ can be found in Ref.~\cite{HHG}.
$E_W$ and $\beta_W$ are the energy and the velocity of 
the $W$ boson in the center-of-mass frame of $\gamma\gamma$ collision, 
respectively.

The cross section for arbitrary initial photon helicities
is given by
\begin{eqnarray}
\widehat{\sigma}_{\gamma\gamma \rightarrow H^\pm W^\mp}^{\lambda_1 \lambda_2}
=\frac{\bar{\beta}}{16\pi s_{\gamma\gamma}}
|{\cal M}^{\lambda_1 \lambda_2}|,
\end{eqnarray}
where $\bar{\beta} \equiv 
\sqrt{1-2\frac{m_{H^\pm}^2+m_W^2}{s_{\gamma\gamma}}+
\frac{(m_{H^\pm}^2-m_W^2)^2}{s_{\gamma\gamma}^2}}$.
The cross section for the process is evaluated by
convoluting with the $\gamma\gamma$ luminosity;
\begin{eqnarray}
\sigma_{ee \rightarrow \gamma\gamma \rightarrow H\pm W\mp}
= \int d\sqrt{s}_{\gamma\gamma} \sum_{\lambda_1 \lambda_2}
 \frac{1}{{\cal L}}
\frac{d{\cal L}^{\lambda_1 \lambda_2}}{d\sqrt{s}_{\gamma\gamma}} 
\widehat{\sigma}^{\lambda_1 \lambda_2}_{\gamma\gamma \rightarrow H^\pm W^\mp}
(\sqrt{s}_{\gamma\gamma}),
\end{eqnarray}
where we use 
the $\gamma\gamma$ luminosity derived from
the tree-level formula of the backward Compton scattering~\cite{gglumi}
for $x=4.8$ assuming complete polarization for laser photons
($P_l=-1.0$) and $90\%$ polarization for electrons ($P_e=0.9$).  
In our numerical evaluation,
the center-of-mass energy of a parent $e^+ e^-$ collider $\sqrt{s}_{ee}$
is assumed to be $1000$ GeV and 
$m_{A^0}=800$ GeV which is the maximum value possible to reach
at $\sqrt{s}_{ee}=1000$ GeV.

\begin{figure*}[h]
\centering
\includegraphics[width=85mm]{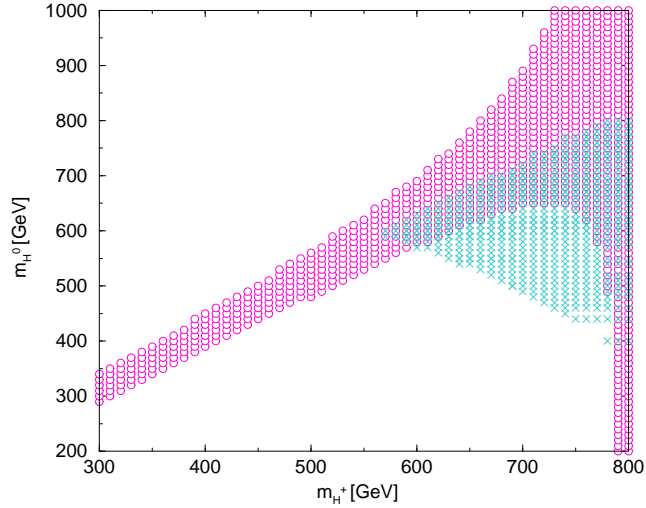}
\caption{The allowed region for $m_{H^\pm}$ and $m_{H^0}$.
The magenta circles (turquoise crosses)
show the allowed region by the rho parameter
measurement (theoretical constraints).
$m_{A^0}=800$ GeV, $\tan\beta=7$, $\alpha=-0.146$.
} \label{rhopunivs}
\end{figure*}

Fig.~\ref{rhopunivs} shows the allowed values of 
$m_{H^\pm}$ and $m_{H^0}$ from various constraints,
for $m_{A^0}=800$ GeV,
$\tan\beta=7$ and $\alpha=-0.146$.
$M$ values are scanned between $0$ and $m_{A^0}$.
While the magenta circles indicate the allowed values by the rho parameter 
measurement~\cite{rho}, the allowed values by theoretical constraints
(the requirment of vacuum stability~\cite{VS} and perturbative
unitarity~\cite{PU} at the tree level) are shown by the turquoise crosses.
As a result, we find that
$m_H^\pm \geq 570$ GeV is allowed, and to assume $m_{H^0}=m_{H^\pm}$
in the numerical evalution is reasonable.

\begin{table}[h]
\begin{center}
\caption{Parameters}
\begin{tabular}{|c|cccc|}
\hline $\tan\beta$ & 1.5 & 7 & 30 & 40
\\
\hline $m_{A^0}$ [GeV] & $~~~800.0~$ & $~~~800.0~$ & $~~~800.0~$ & $~~~800.0~$
\\
\hline $m_{H^0}$ [GeV] & $~~m_{H^\pm}$ & $~~m_{H^\pm}$ 
& $~~m_{H^\pm}$ & $~~m_{H^\pm}$
\\
\hline $m_{h^0}$ [GeV] & $~~~97.7~$ & $~~~119.6~$ & $~~~121.4~$ & $~~~121.54~$ 
\\
\hline $~M~$ [GeV] & $~0.0-m_{A^0}$ & $~0.0-m_{A^0}$ 
& $~0.0-m_{A^0}$ & $~0.0-m_{A^0}$ 
\\
\hline $\alpha$ & $-0.600$ & $-0.146$ & $-0.0337$ & $-0.0251$ 
\\
\hline
\end{tabular}
\label{parameters}
\end{center}
\end{table}

\begin{figure*}[t]
\centering
\includegraphics[width=85mm]{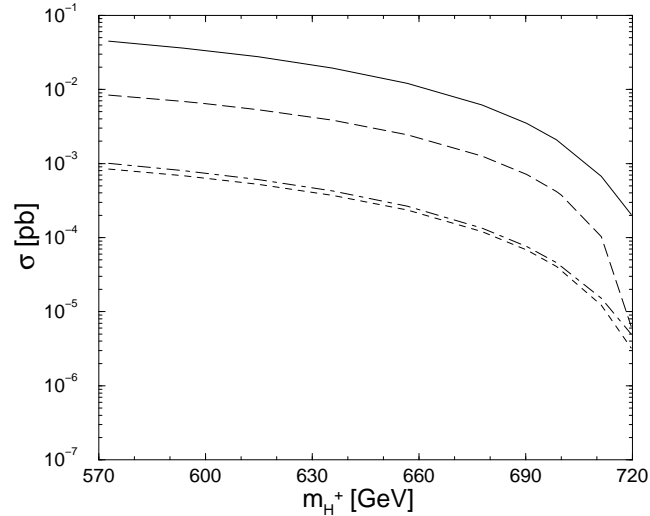}
\caption{The cross sections for $\gamma \gamma \rightarrow A^0
\rightarrow H^\pm W^\mp$. $m_{A^0}=800$ GeV, $\sqrt{s}_{ee} = 1000$ GeV.
The solid (dashed, dotted and dat-dashed) curve 
indicates the case for $\tan\beta=1.5$ (7, 30 and 40).
} \label{crosssection}
\end{figure*}

The cross sections for four reference choices of $\tan\beta$
($\tan\beta=1.5$, $7$, $30$ and $40$) are shown in Fig.~\ref{crosssection}.
The parameters used for evaluating
each case are listed in Table~\ref{parameters}.
The $M$ values allowed by the above constraints are 
dependent on $m_{H^0}$. In the numerical evalution, the relevant
values of $M$ between $0$ to $m_{A^0}$ are used. 
The dependence of the cross sections on the choices of the $M$ values
is insignificantly small.

Though the cross sections for the $\gamma\gamma \rightarrow H^\pm W^\mp$
process are less than $0.1$ fb in the MSSM~\cite{MK}, the cross sections
for the $\gamma\gamma \rightarrow A^0 \rightarrow H^\pm W^\mp$
process amount to $0.1 - 100$ fb in general. Such cross section
values are comparable with the ones for other
charged Higgs boson production processes at a photon collider~\cite{MK}.
While for other processes the cross sections are too small
for $m_H^\pm \geq 500$ GeV, $0.1 - 100$ fb can be expected
in the $\gamma\gamma \rightarrow A^0 \rightarrow H^\pm W^\mp$
process for $m_H^\pm \geq 570$ GeV. 

\section{SUMMARY}

In the THDM, the decay of a Higgs boson into another Higgs boson
can be allowed
within experimental and theoretical constraints,
though this type of decay
is hard to occur in the decoupling region of the MSSM.
To study this type of decay
may bring us an opportunity to distinguish between the MSSM
and such a model~\cite{ABK}, as well as to affect
detection strategies for Higgs bosons.

The $\gamma \gamma \rightarrow A^0 \rightarrow H^\pm W^\mp$ 
processes have been considered here. 
The $A^0$ resonant production in the processes can be realized
in the THDM.
For $\sqrt{s}_{ee}=1000$ GeV, it is possible to produce the
$A^0$ bosons whose mass is less than
$800$ GeV at a photon collider.
We have studied the case where $\sqrt{s}_{ee}=1000$ GeV and
$m_{A^0}=800$ GeV.

Though the cross sections for the $\gamma\gamma \rightarrow H^\pm W^\mp$
processes are less than $0.1$ fb in the MSSM,
the cross sections for the
$\gamma \gamma \rightarrow A^0 \rightarrow H^\pm W^\mp$ processes
in the THDM
can reach $0.1 - 100$ fb. It is important that 
such cross section values can be realized
for $m_H^\pm \geq 570$ GeV where the values for other
charged Higgs boson production processes are too small.
Therefore, even
if the charged Higgs bosons are too heavy to be detected 
in other charged Higgs boson
production processes at the LC,
it may be possible to detect them in this process.

\end{document}